\documentclass[prl,twocolumn]{revtex4}

\usepackage{epsfig}
\usepackage{dcolumn}
\usepackage{bm}
\begin{document}

\title{OPTIMAL PUMP FREQUENCY
FOR AC HYSTERETIC SQUID}
\author{Andrey L. Pankratov }
\email{alp@ipm.sci-nnov.ru}
\affiliation{Institute for Physics of Microstructures of RAS,
Nizhny Novgorod, RUSSIA.}

\begin{abstract}
This paper presents the analytical and numerical analysis of
fluctuational dynamics of ac hysteretic SQUID. It has been
demonstrated, that the most important parameter for the improvement
of the hysteretic SQUID sensitivity is the
ratio between the pump and the characteristic frequencies but not
their absolute values. The resonant behavior of signal-to-noise
ratio as function of the pump frequency has been observed and the
narrow area for the optimal pump frequency range is indicated.
\end{abstract}

\maketitle

It is well known \cite{Likharev},\cite{Barone},\cite{Clark2}
that the slope of a signal
characteristic of hysteretic ac SQUID
$h=dV_{ac}/d{\mit\Phi}$ increases with increasing the
working (pump) frequency and therefore output energy sensitivity
becomes considerably better.
Noise properties of ac SQUIDs were intensively investigated both
for cases $\Omega\ll 1$ and $\Omega\gg 1$ (where $\Omega=\omega/
\omega_c$, $\omega$ is the pump frequency and $\omega_c$ is the
characteristic frequency of Josephson junction (JJ))
\cite{Kur},\cite{Dan},\cite{Snig}.
It has been demonstrated  that increasing of frequency
without violation of working capacity
of SQUID is possible nearly up to the characteristic frequency of JJ
\cite{BJ}, which may be of the order of 1 THz for JJ based
on low $T_{c}$ materials and 10 THz - for high
$T_{c}$ ones. Besides, it was reported \cite{Long} about ac SQUID
with $\Omega=0.1$, demonstrated high sensitivity about $4\cdot
10^{-32}$J/Hz.
So, elaboration and fabrication of a single
junction SQUID working in microwave frequency range looks very
promising, and usual opinion is that increasing of the pump frequency may
allow to reach, in principle, such high characteristics as dc SQUIDs
have.

However, at the present time, there is still no
at least qualitative understanding at which pump frequency a
microwave SQUID should operate to reach the maximal sensitivity, since
due to mathematical problems no investigations were performed in the most
interesting pump frequency range $\omega\sim\omega_c$.
As follows from recent results by Cheska \cite{ch}, there should not
be such optimal pump frequency for a nonhysteretic SQUID,
which is not surprising since that is a (quasi) linear system.
The ac hysteretic SQUID, is, however, strongly nonlinear system,
and, recently, in qualitatively similar model system, the resonant behavior
of signal-to-noise ratio as function of frequency of driving signal
has been observed \cite{PRE}.

The present paper is devoted to studying the nonlinear
fluctuational dynamics of a hysteretic ac SQUID in
the pump frequency range around $\omega_c$. The effect of suppression of noise
by strong periodic signal \cite{PRE} has for the first time been observed
in the model, describing the real electronic device:
it has been demonstrated that there exists a certain optimal
pump frequency range at which the hysteretic SQUID will
operate with minimal noise-induced error and that there
is an optimal driving amplitude at which the voltage-flux
characteristic is the most close to the noiseless case.

Theory of a single junction SQUID is already quite well developed (see,
e.g., \cite{Likharev}-\cite{Clark2} and
references therein).
Necessary requirements for SQUIDs operating in hysteretic mode are the
following: $\ell \geq 1$ to provide hysteresis and
maximize nonlinearity (where $\ell=L/L_0$,
$L$ is the inductance of the ring,
$L_0={\mit\Phi}_0/(2\pi I_c)$, $I_c$ is the critical
current of the JJ, ${\mit\Phi}_0$
is the flux quantum) and
$\beta=2\pi I_cR_N^2C/{\mit\Phi}_0\ll 1$
(the SQUID should be overdamped) to
prevent from stochasticity when switching between states with
different trapped fluxes in the SQUID ring (here
$R_N^{-1}=G_N$ is the normal conductivity of the JJ, $C$ is the
capacitance).

The fluctuational dynamics of a flux in a SQUID ring
coupled to a resonator may be described by
the following equations:
\begin{eqnarray}
\dot{\varphi}+ {\ell}\sin \varphi +\varphi-\varphi_{m}-\alpha_s\psi(t)
=\varphi_F(t),  \label{ring}
\end{eqnarray}
\begin{eqnarray}
\frac{1}{\omega_0^2}\ddot{\psi}+\frac{\eta_r}{\omega_0}\dot{\psi}
+\psi=\alpha_r(\varphi-\varphi_{m})+a\sin(\omega_0 t).
\label{res}
\end{eqnarray}
Here $\varphi=\displaystyle\frac{2\pi{\mit\Phi}}{{\mit\Phi}_0}$,
${\mit\Phi}$ is the trapped flux,
$\varphi_{m}=\displaystyle\frac{2\pi{\mit\Phi_m}}{{\mit\Phi}_0}$,
${\mit\Phi_m}$ is the measured flux and $\psi(t)$
is the pump signal from the resonator,
time is normalized to the characteristic frequency
of the SQUID $\omega_s=\omega_{c}/\ell=R_N/L$,
$\omega_{c}=
2\pi R_NI_c/{\mit\Phi}_0$ -- characteristic
frequency of JJ, $\eta_r$ is the damping of
the resonator, $\alpha_s$ and $\alpha_r$ are coupling
coefficients of the SQUID ring and the resonator, respectively,
and $\omega_0=\omega_r/\omega_s$ is the dimensionless pump
frequency.

Let us take into account only internal thermal fluctuations
in the SQUID ring. In this case the noise source $\varphi_F(t)$ may be
represented by the white Gaussian noise:
\begin{eqnarray}
\left<\varphi_F(t)\right>=0,\quad
\displaystyle{\left<\varphi_F(t)\varphi_F(t+\tau )\right>=
{2\gamma \ell}\delta (\tau )},
\label{gamma}
\end{eqnarray}
where $\gamma=2\pi k_B T/({\mit\Phi}_0 I_c)$
is the dimensionless noise intensity, $T$ is the temperature and
$k_B$ is the Boltzmann constant.

    Let us first consider analytically and
numerically the fluctuational flux dynamics of a SQUID ring,
supposing that the signal from the resonator is known:
$\alpha_s\psi(t)=A\sin(\omega_0 t+\xi)$, where $A$ is the driving amplitude
and $\xi$ is the initial phase.
    In this case the required statistical characteristics
may be computed on the basis of the transitional
probability density $W(\varphi,t)$.
It is well known, that the Fokker-Planck equation (FPE) for the probability
density $W(\varphi,t)$ corresponds to Eq. (\ref{ring}) for the flux:
\begin{eqnarray}\nonumber
{\partial W(\varphi,t)\over\partial t}=
-\frac{\partial G(\varphi,t)}{\partial \varphi}= \\
={\partial\over\partial \varphi}\left\{\left[
{du(\varphi,t)\over d\varphi}W(\varphi,t)\right]+
\gamma\ell{\partial W(\varphi,t)\over\partial
\varphi}\right\},
\label{FPE}
\end{eqnarray}
where
\begin{equation}\label{pot}
u(\varphi )=-\ell\cos(\varphi)+(\varphi-\varphi_{e})^2/2
\end{equation}
is the dimensionless potential profile,
$\varphi_{e}=\varphi_{m}+A\sin(\omega_0 t+\xi)$ is the external
flux.
The initial and boundary conditions for Eq. (\ref{FPE}) are as
follows:
\begin{center}
    $W(\varphi,0)= \delta(\varphi-\varphi_0)$ and $G(\pm\infty,t)=0.$
\end{center}

The quantity, that reflects the noise induced transition process
from the metastable state is a probability
$\displaystyle{P(t)=\int_{-\infty}^d
W(\varphi,t)d\varphi}$ that transition will not occur at the
moment of time $t$, that we will call the survival probability,
where $d$ is some boundary point, usually a potential barrier
top.

To describe analytically the survival probability one can
recourse to the adiabatic approximation, that
for the case of a weak periodic driving has been used in the
context of stochastic resonance
\cite{MW},\cite{st}. It has been demonstrated in \cite{PLA}
(see also \cite{ACP}) that for the case of a metastable potential
and strong periodic driving the evolution of survival probability
in the frequency range $0\le \omega_0< 0.5$ may be well described by a
modified adiabatic approximation, which allows to extend the
usual analysis to arbitrary driving amplitudes and noise
intensities. For simplicity, let us consider the case for
$\varphi_m=\pi$ (bistable potential), then the survival probability
may be written in the form:
\begin{eqnarray}\label{met}
P(\varphi_0,t)=g(t)\left[P(\varphi_0,0)+
\int_{0}^t
\frac{dt'}{g(t')\tau_q(\varphi_0,t')}\right],\\
g(t)=\exp\left\{-\int_{0}^t
\left[\frac{1}{\tau_p(\varphi_0,t')}+\frac{1}{\tau_q(\varphi_0,t')}
\right]dt'\right\},\nonumber
\end{eqnarray}
where $\tau_p(\varphi_0,t')$ is the exact mean decay time
of the left metastable state obtained for the corresponding
time-constant potential \cite{M},\cite{ACP}:
\begin{eqnarray}\label{tp}
\tau_p(\varphi_0) = \frac{1}{\gamma\ell}\left\{\int\limits_{\varphi_0}^d
e^{u(y)/\gamma\ell}\int\limits_{-\infty}^y e^{-u(x)/\gamma\ell}dxdy \right. +\\
+\left. \int\limits_{d}^{c} e^{u(y)/\gamma\ell}dy
\int\limits_{-\infty}^d e^{-u(x)/\gamma\ell}dx \right\}. \nonumber
\end{eqnarray}
Here $c>d$ is the coordinate of the right potential minimum
at the moment when $\sin(\omega_0 t+\xi)=1$ and the left minimum
disappears. The time scale $\tau_q$ is the decay time of the
right metastable state and is calculated from the same formula
(\ref{tp}) where in the potential (\ref{pot}) the phase of the
driving signal is changed from 0 to $\pi$.
Note that with respect to the usual adiabatic analysis the
approximate Kramers' time has been substituted by the exact one
(\ref{tp}) and a surprisingly good agreement of this approximate
expression with the computer simulation results has been found in
rather broad range of parameters, that is seen in Fig. 1.

\begin{center}
\epsfxsize=7.5cm \epsfbox{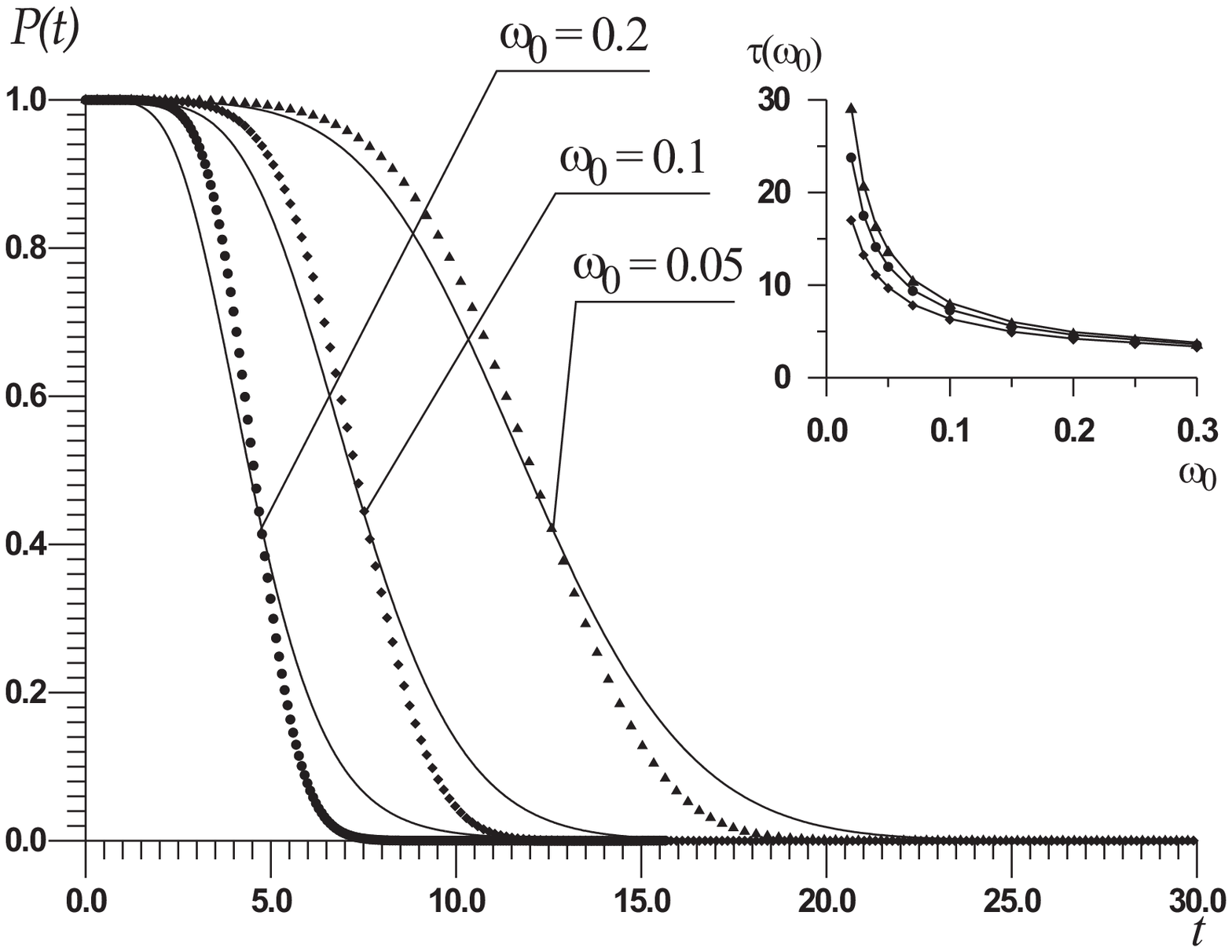}
\small{Fig. 1. The evolution of survival probability; dots - computer simulations,
solid line - modified adiabatic approximation. Inset: the mean transition time;
dots - computer simulations, solid line - modified adiabatic approximation;
$\gamma=0.04;0.1;0.2$ from top to bottom.}
\end{center}

In Fig. 1 the probability evolution is presented for $\gamma=0.1$,
$A=3$, $\xi=0$ and $\varphi_m=\pi$. The adiabatic
approximation is drawn by solid line, while the results of
numerical solution of Eq. (\ref{FPE}) by dots. One can see, that
the coincidence is good enough.
Even better agreement between the modified adiabatic
approximation (\ref{met}), (\ref{tp}) and the computer
simulation results may be observed for
the mean decay time of the left metastable state (inset of Fig. 1).
Since the working conditions for the considered potential
are such that during half of the driving period $T_{\omega}$
the survival probability monotonically changes from unity to
almost zero, we can define the mean decay time in time-periodic potential as:
\begin{eqnarray}\label{tau}
\tau_p(\omega_0) &=& \frac{\int_0^{T_{\omega}/2} [P(t)-P(0)]dt}
{P(T_{\omega}/2)-P(0)}.
\end{eqnarray}
The definition (\ref{tau}) is analogous to widely used definition
of integral relaxation time for diffusion in
time-constant potentials (see \cite{ACP} and references therein).
As it is seen from the inset of Fig. 1, in the range of validity of adiabatic
approximation the mean decay time (\ref{tau}) computed using
formulas (\ref{met}),(\ref{tp}) gives perfect coincidence with
the results of computer simulation.
It is intriguing to see that there is a range of
driving frequencies $0.1\le\omega_0\le 0.3$, where the mean decay
time is almost insensitive to the noise intensity: the curves for
different noise intensities actually coincide that constitutes
the effect of suppression of noise by a strong periodic signal.
This effect may also be demonstrated by the analysis of the
signal-to-noise ratio $R$. In accordance with \cite{st}
we denote $R$ as:
\begin{equation}\label{snr}
R=\frac{1}{S_N(\omega_0)} \lim_{\Delta \omega \to 0}
\int\limits_{\omega_0-\Delta\omega}^{\omega_0+\Delta\omega}
S(\omega) d\omega,
\end{equation}
where
$S(\omega)=\int_{-\infty}^{+\infty} e^{-i\omega \tau}
K[\tau]d\tau$
is the spectral density, $S_N(\omega_0)$ is noisy pedestal at
the driving frequency $\omega_0$ and $K[\tau]$ is the correlation
function:
$K[\tau]=\left< \left< \varphi(t+\tau)\varphi(t)\right> \right>$,
where the inner brackets denote the ensemble average and outer
brackets indicate the average over initial phase $\xi$.
In order to obtain $K[\tau]$ the
Eq. (\ref{FPE}) has been solved numerically for $A=3$,
$\ell=3$, $\varphi_m=\pi$.

\begin{center}
\epsfxsize=7.5cm \epsfbox{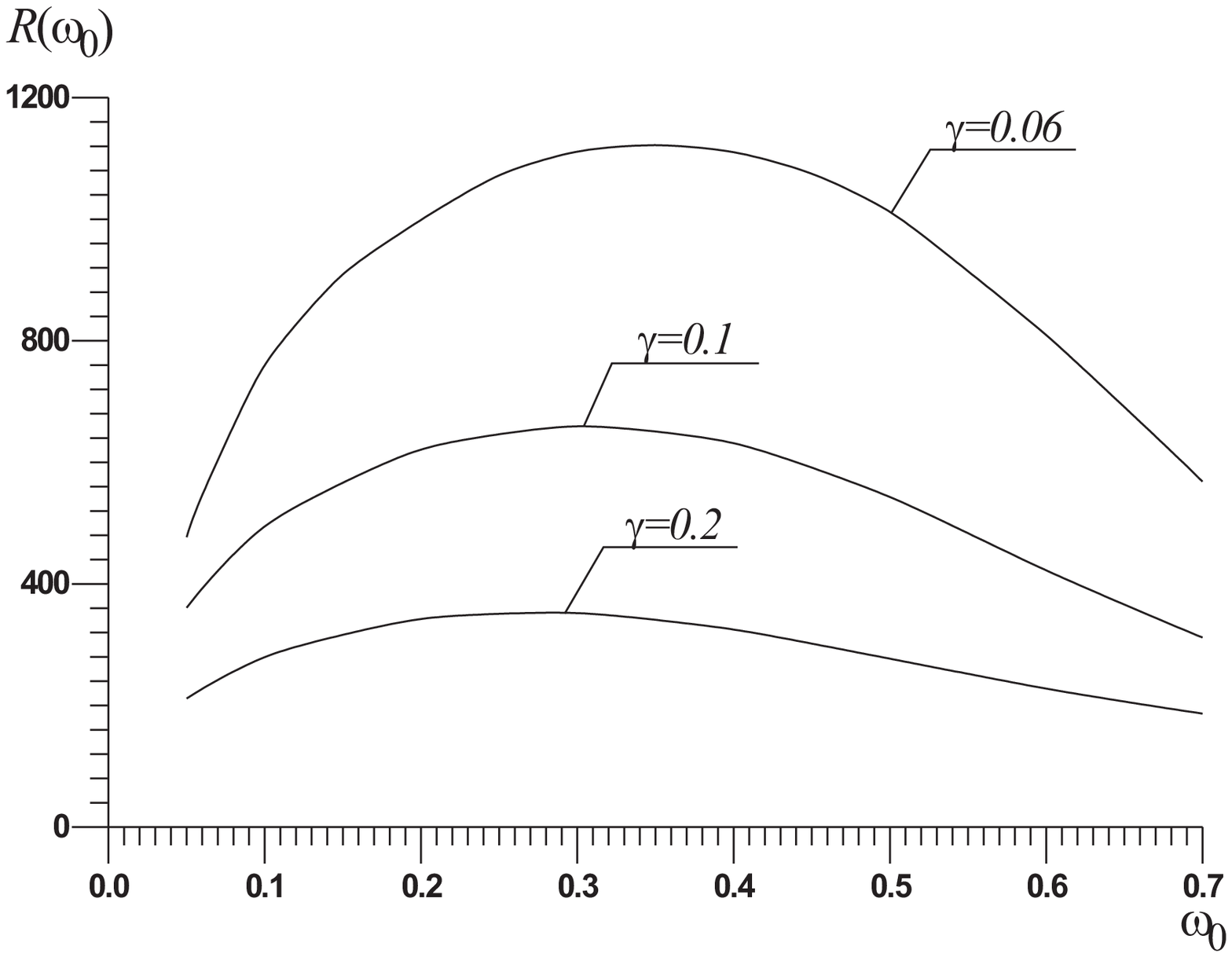}
\small{Fig. 2. The signal-to-noise ratio as function of driving frequency.}
\end{center}

Let us plot the signal-to-noise ratio (SNR)
as function of driving frequency $\omega_0$. From Fig. 2 one can
see, that $R$ as function of $\omega_0$ has strongly pronounced
maximum. The location of this maximum lies below the cut-off
frequency ($\omega_0=1$ or $\omega_s=R_N/L$
in dimensional notations) and depends on the driving amplitude
$A$ and noise intensity. The existence of
optimal driving frequency may be explained
in the following way: with increase of frequency noise-induced escapes
occur at lower barrier, and the maximum of SNR corresponds to the lowest barrier,
that is closest to the dynamical case, where the transition occur when the barrier
completely disappears. With further increase of frequency, there is not enough time
for transition in the absence of noise and SNR drops, while the mean transition time
rises \cite{PRE},\cite{PLA}.

One can solve numerically the system of equations
(\ref{ring}) and (\ref{res}) and compute ac current-voltage
characteristic and voltage-flux characteristic.

\begin{center}
\epsfxsize=7.5cm \epsfbox{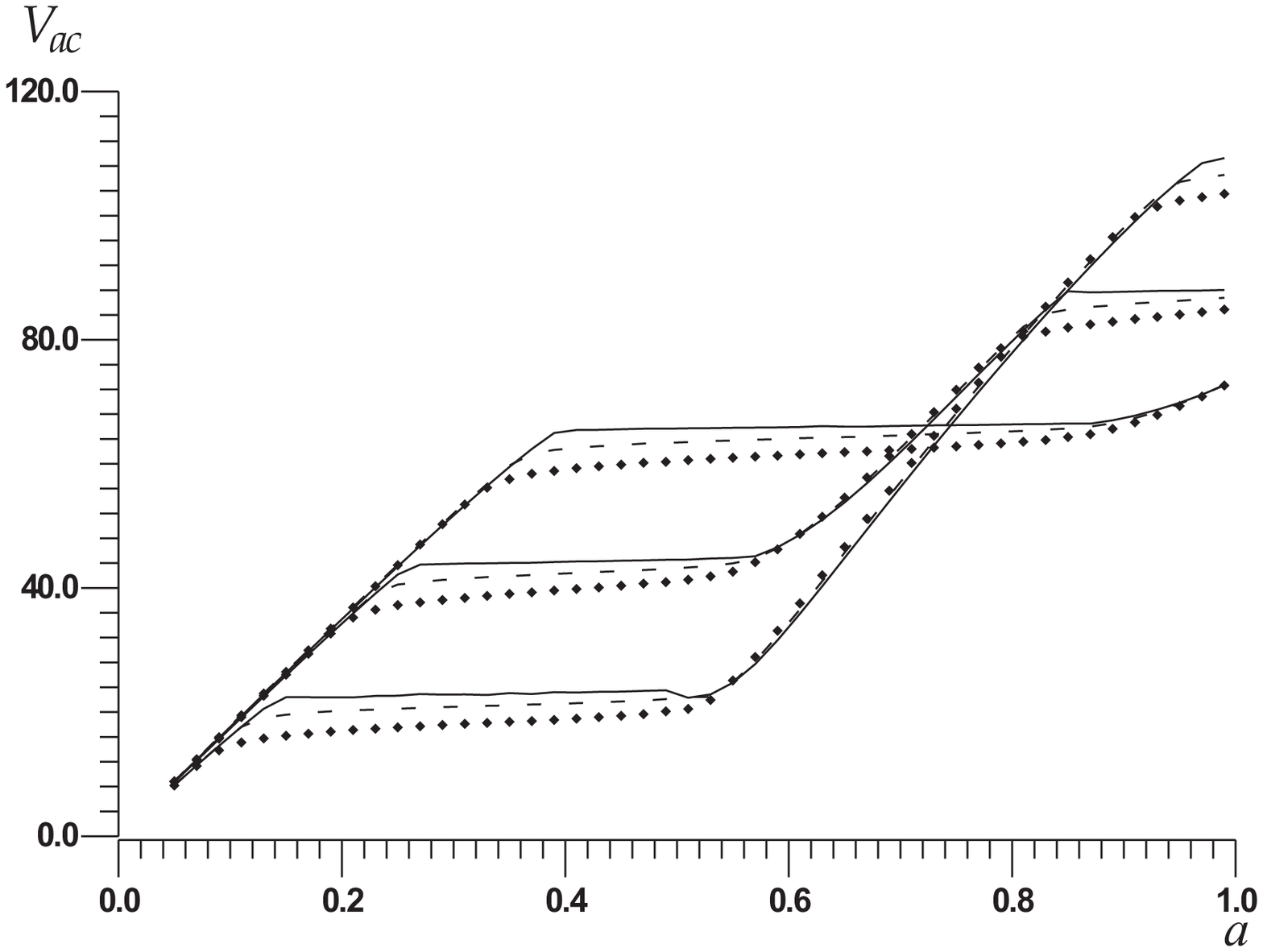}
\small{Fig. 3. The ac current-voltage characteristic for $\omega_0=0.01$;
solid line - $\gamma=0$, dashed line - $\gamma=0.01$, diamonds -
$\gamma=0.03$. }
\end{center}
\begin{center}
\epsfxsize=7.5cm \epsfbox{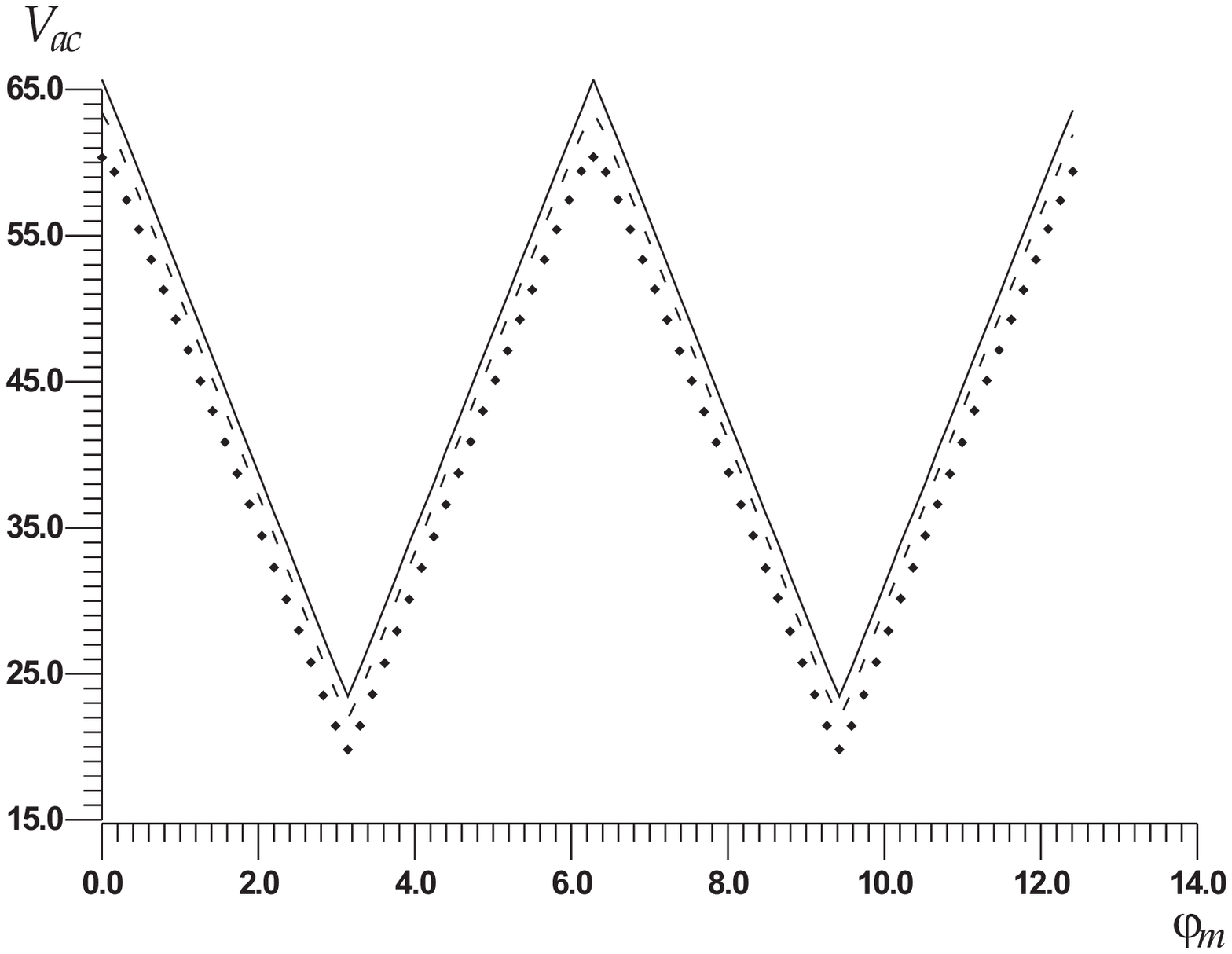}
\small{Fig. 4. The voltage-flux characteristic for $\omega_0=0.01$;
solid line - $\gamma=0$, dashed line - $\gamma=0.01$, diamonds -
$\gamma=0.03$.}
\end{center}

In Fig. 3 the ac current-voltage characteristic ($V_{ac}(a)=\sqrt{2\overline{\psi^2}}$,
$\varphi_m=0;\pi/2;\pi$) is presented for $\omega_0=0.01$ that corresponds to relatively low frequency
case and for the following parameters: $\ell=3$, $\eta_r=0.01$, $\alpha_s\alpha_r/\eta_r=2$.
As it is seen, for the case $\gamma=0$ the plateaus are
nearly horizontal. For $\gamma=0.01$ and $\gamma=0.03$ the plateaus
have a tilt and lie below the curves for $\gamma=0$.
The voltage-flux characteristics for the same parameters and $a=0.47$
is presented in Fig. 4. It is seen that with increase of noise intensity
$\gamma$ the corresponding curves move down and their edges are rounded.
Low frequency interference, as well as noise of the resonator and amplifiers
(that were not taken into account in the present paper) will make the situation only
worse and the quantity of splitting of the curves for different $\gamma$ will give
the error of the measured flux $\varphi_m$. Fortunately, with the increase of the pump
frequency the picture changes radically: already at $\omega_0=0.1$ the plateaus for
$\gamma=0.01$ cross the plateaus for $\gamma=0$ at certain points. With approaching
the maximum of signal-to-noise ratio $\omega_0\approx 0.3$ (Fig. 2) the crossing
points move to the middle of the plateau and for different $\gamma$ become more
close to each other, see Fig. 5.

\begin{center}
\epsfxsize=7.5cm \epsfbox{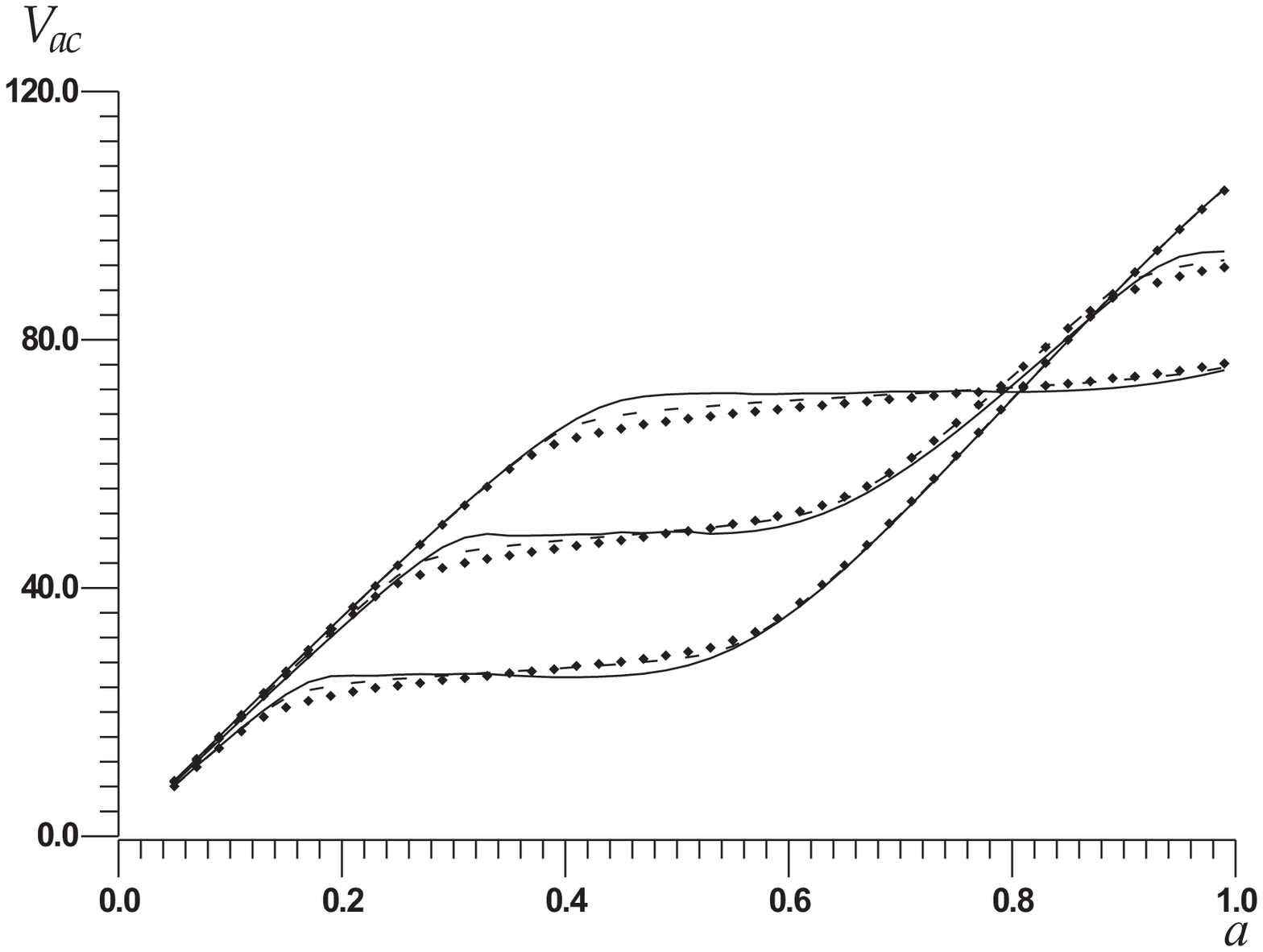}
\small{Fig. 5. The ac current-voltage characteristic for $\omega_0=0.3$;
solid line - $\gamma=0$, dashed line - $\gamma=0.01$, diamonds - $\gamma=0.03$.}
\end{center}
\begin{center}
\epsfxsize=7.5cm \epsfbox{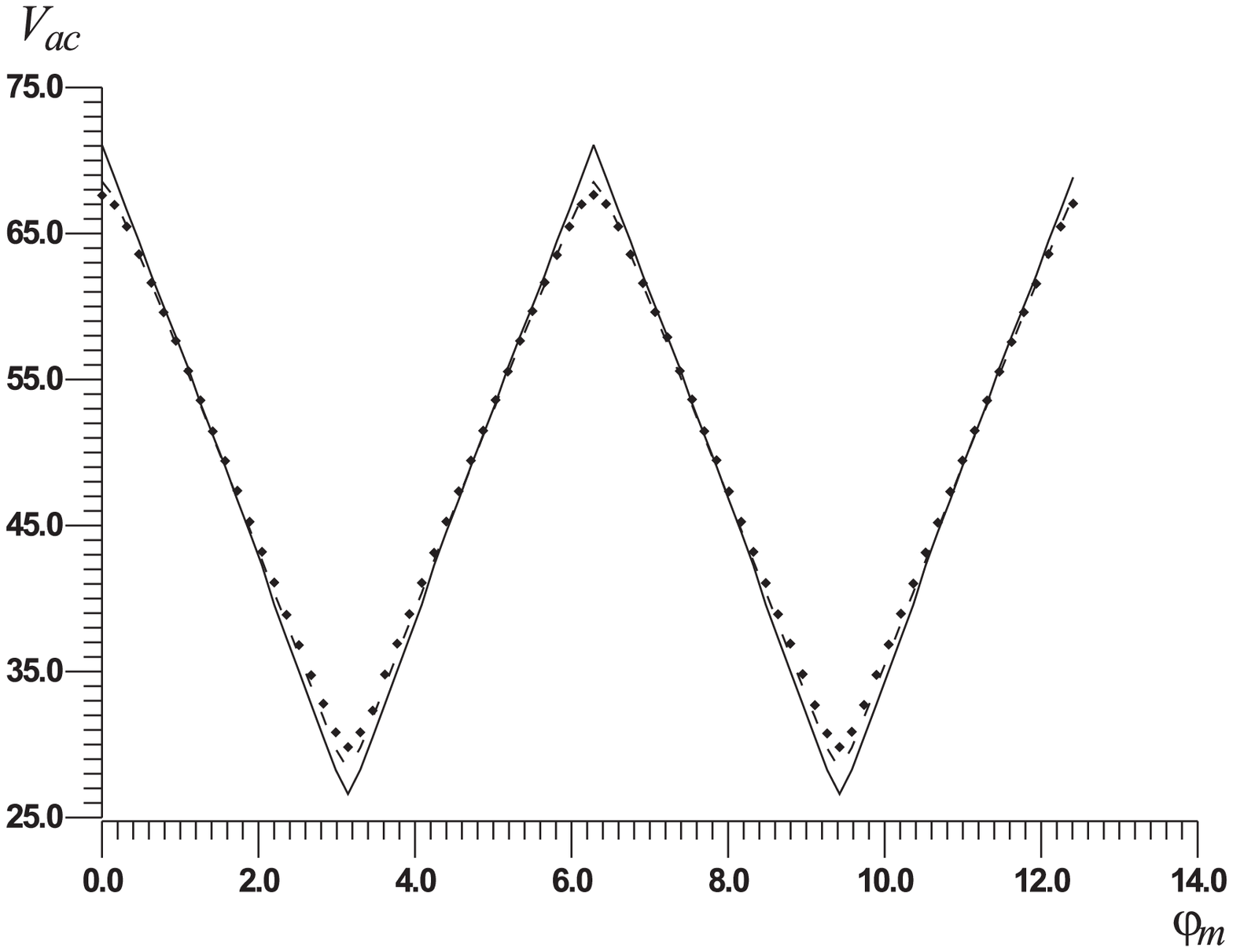}
\small{Fig. 6. The voltage-flux characteristic for $\omega_0=0.3$;
solid line - $\gamma=0$, dashed line - $\gamma=0.01$, diamonds - $\gamma=0.03$.}
\end{center}

It is intuitively obvious that to reach the maximal sensitivity of the SQUID,
the voltage-flux characteristic should be computed for the pump amplitude $a$, corresponding to
the crossing point of plateaus for $\varphi_m=\pi/2$ (see analogous consideration
for dc SQUIDs in \cite{Prok}). In Fig. 6 the voltage-flux characteristic is plotted
for $\omega_0=0.3$, $a=0.51$ and $\gamma=0;0.01;0.03$. It is seen, that there is a
broad range of $\varphi_m$, $0.7\le \varphi_m\le 2.2$, where curves for different $\gamma$ coincide
and, therefore, noise actually do not affect the SQUID in this range of parameters.
If, however, the measured flux lie outside the range $0.7\le \varphi_m\le 2.2$, some
known quantity of magnetic flux $\varphi_n$ may be added to shift the working point to this
region, precise measurements may be performed, and $\varphi_n$ may be extracted to get
information about the original measured flux.

Up to date tendency in ac SQUID development is increasing the
pump frequency up to microwave range ($\sim 10GHz$) in order to
reduce the noise \cite{Clark2} with respect to
measured voltage and thereby to increase the sensitivity of
devices of this kind.
However, as follows from the above
presented analysis,
the main optimizational parameter of the microwave
hysteretic SQUID is the ratio between the pump and the characteristic frequencies
$\omega_0=\omega_r\ell/\omega_c$, but not their absolute values;
it must be in the range $\omega_0\approx 0.2-0.4$ in order to get maximal
sensitivity of the SQUID. Also during measurements, very important parameter is
the amplitude of ac driving. It can be chosen from
ac current-voltage characteristic of the SQUID as the
point, where the plateau for ${\mit\Phi}_m={\mit\Phi}_0/4$ for a given temperature $k_BT\ne 0$
crosses the plateau for $k_BT=0$; the calibration can be done once
by cooling the SQUID to a low temperature.

The author wishes to thank V. V. Kurin and Yu. N. Nozdrin
for helpful discussions. The work has been supported by the Russian
Foundation for Basic Research (projects 00-02-16528, 02-02-16775,
02-02-17517, 02-02-06126, and 00-15-96620) and by INTAS (projects
01-0367 and 01-0450).

\end{document}